\documentclass[journal,twocolumn]{IEEEtran}
\usepackage{amssymb}
\usepackage{amsfonts}
\usepackage{amsmath}
\usepackage{algorithm}
\usepackage{algorithmic}
\usepackage{subeqnarray}
\usepackage{cases}

\usepackage{graphicx,subfigure,amsmath,amssymb,cite}

\usepackage[dvips]{color}
\usepackage{float}
\ifCLASSINFOpdf
\else
\fi
\begin{document}

\title{Two Practical Random-Subcarrier-Selection Methods for Secure Precise  Wireless Transmission}

\author{Tong Shen,~Shuo Zhang,~Riqing Chen,~Jin Wang,~Jinsong Hu,~Feng Shu,~and Jiangzhou Wang

\thanks{This work was supported in part by the National Natural Science Foundation of China (Nos. 61771244, 61501238, 61702258, 61472190, and 61271230), in part by the Open Research Fund of National Key Laboratory of Electromagnetic Environment, China Research Institute of Radiowave Propagation (No. 201500013), in part by the Jiangsu Provincial Science Foundation under Project BK20150786, in part by the Specially Appointed Professor Program in Jiangsu Province, 2015, in part by the Fundamental Research Funds for the Central Universities under Grant 30916011205, and in part by the open research fund of National Mobile Communications Research Laboratory, Southeast University, China (Nos. 2017D04 and 2013D02)(Corresponding author: Feng Shu).}
\thanks{Tong Shen,~Jin Wang,~and Feng Shu are with the School of Electronic and Optical Engineering, Nanjing University of Science and Technology, 210094, CHINA. (Email: shufeng0101 @163.com). }
\thanks{Feng Shu, ~Jin Wang,~ and Riqing~Chen are  with the School of Computer and  information at Fujian Agriculture and Forestry University, Fuzhou, 350002, China.}
\thanks{Shuo Zhang is with the National Key Laboratory of Science and Technology on Aerospace Intelligence Control, Beijing Aerospace Automatic Control Institute, Beijing, 100854, China. (Email: gcshuo@163.com)}
\thanks{Jiangzhou Wang is with the School of Engineering and Digital Arts, University of Kent, Canterbury CT2 7NT, U.K. Email: \{j.z.wang\}@kent.ac.uk.
}


}

\maketitle

\begin{abstract}
In  directional modulation (DM) networks, two practical random-subcarrier-selection (RSS) methods are proposed to transmit confidential message to the desired user per orthogonal frequency division multiplexing (OFDM) symbol with only single receive power peak formed by constructing random subcarrier set and performing a randomization procedure. This scheme completely addresses the crucial problem facing secure precise wireless transmission (SPWT), how to achieve the SPWT per OFDM symbol while the traditional SPWT holds only in the statistically average sense. Several necessary conditions for SPWT per OFDM is derived and proposed: random, distributed, and distinct subcarrier spacing pattern mapped onto transmit antenna array. Two random subcarrier index sets, quadratic subcarrier set (QSS) and prime subcarrier set (PSS), are constructed. Subsequently,  a random  metric is defined, and a randomization procedure (RP) is proposed. Its detailed process includes the following steps: integer mod, ordering, blocking, and block interleaving (BI) where BI is repeated until the random metric is greater than the predefined value. This yields a single high  energy main peak (SHEMP) at the desired position with other positions, outside the main peak, harvesting only weak receive energy seriously corrupted by AN. Simulation results confirm that our proposed two methods, QSS plus RP and PSS plus RP,  perform significantly better than conventional methods such as linear subcarrier set (LSS) plus RP. They actually form a SHEMP at desired position, and generate only extremely small leakage outside the main peak.
\end{abstract}
\begin{IEEEkeywords}
Secure precise wireless transmission, direction modulation, random subcarrier-selection, quadratic subcarrier set, prime subcarrier set, randomization procedure
\end{IEEEkeywords}
\section{Introduction}
 Physical-layer security (PLS) in wireless networks is an extremely important and challenging research field and has made a great progress in the last decade \cite{Guo2017Exploiting,Yan2016Artificial,Ma2017Interference,Zou2016Relay,Alodeh2017Symbol,Zhou2018Computation,Trappe2015The,Yan2016Location,Chen2016A,Chen2015Large,Wang2016Physical,Wang2015Secure,Zou2014Secrecy}. As an efficient way for PLS wireless transmission, directional modulation (DM) has attracted extensive research interesting and activities \cite{Daly2009Directional,Yuan2014A,Hu2016Robust,Shu2017Robust,Shu2017Artificial,Wan2018Power,Shu2018Directional}. In  \cite{Goel2008Guaranteeing, Negi2005Secret}, the authors first proposed to use artificial noise (AN) projection to degrade the performance of eavesdropper with the help of multiple transmit antennas. This opens a new way of PLS in wireless networks, which also triggered the rapid development of PLS. In \cite{Daly2009Directional}, the authors proposes a DM array actively driven by utilizing analog radio frequency phase shifters or switchers, but this scheme has its weakness that it is difficult to design such a high speed RF switchers or shifters and the complexity of its design process is too high. To address this difficulty, the authors in \cite{Yuan2014A} proposed a DM synthesis scheme on baseband with the help of orthogonal AN. In \cite{Hu2016Robust,Shu2017Robust}, the authors took direction measurement error into account and propose two robust DM synthesis schemes with AN in single desired user and multi-user scenarios respectively. This scheme can significantly improve the bit error rate (BER) performance in desired receivers compared with non-robust methods. In a multi-user DM situation, without the prior knowledge of measurement error distribution of directional angles, a main-lobe-integration-based leakage beamformer is proposed to realize the robust secure wireless transmission\cite{Shu2017Robusts}. In multi-cast scenario, a secure pre-coding method was proposed in \cite{Shu2017Artificial} with the aid of AN. In \cite{Wan2018Power}, given any precoder of confidential messages and AN projection matrix, the authors proposed an optimal power allocation strategy of maximizing the secrecy rate in DM networks. With this strategy, the secrecy rate has been greatly improved.
 
 DM is a key-less and important transmit way of PLS, but there also exists a serious secure problem. That is, the beamforming method of DM only depends heavily on direction but is independent of distance. It has the possibility that the eavesdroppers move inside the main beam of the desired user, implying that eavesdropper can readily intercept the confidential messages from the channel.  In such a scenario, the DM can no longer guarantee the secure transmission of confidential messages.

To overcome such a problem, in \cite{Hu2017SPWT}, the authors combined random frequency diverse array (RFDA) in \cite{Sammartino2013Frequency, Wang2015Frequency,Zhu2012Radio} and direction modulation in \cite{Daly2009Directional,Yuan2014A} to propose an excellent secure transmission idea, called secure precise wireless transmission (SPWT). In \cite{Zhu2017Secure}, the authors also proposed to achieve a SPWT with the help of cooperative relays as a second way. The SPWT is viewed as an emerging technology based on direction modulation (DM). Its basis idea is to form a single high energy main peak (SHEMP)  at the desired position. If eavesdropper moves outside the peak, it only intercepts the weak receive confidential signal severely corrupted by AN, and thus cannot detect the confidential messages correctly. However, for a SPWT of using RFDA plus DM, as the number $N$ of antenna-array elements tends to be medium-scale or large-scale, the receiver requires $N$-RF chains to coherently combine the transmit signals. In other words, the circuit complexity for the desired receiver will become prohibitive. Therefore, in \cite{Shu2018SPWT}, the authors proposed a SPWT transmit structure using random subcarrier selection (RSS) of orthogonal frequency division multiplexing (OFDM) instead of RFDA with a low-complexity receiver structure and orthogonal property among random frequencies. This FFT/IFFT operation structure will significantly simplify the circuit complexity and cost of coherent combination at receiver.

However, in both \cite{Shu2018SPWT} and \cite{Hu2017SPWT}, they only proposed an elementary idea: SPWT. Actually, in the two papers, a single energy peak is formed at the desired position in the statistically average sense, not a practical transmit way. How to achieve a SPWT per OFDM symbol?  That is, only single high energy peak is produced at receive terminal per OFDM symbol.  This motivates us to propose some practical SPWT procedures. In this paper, we focus mainly on the investigation of constructing some practical SPWT schemes. Our main contributions are summarized as follows:

\begin{enumerate}
\item Given a matched filtering precoder, which achieves a phase alignment at desired position, to observe all phases at undesired receiver, we find a fact that when $N_T$ receive signal phases from $N_T$ transmit antennas are uniformly and sparsely distributed over unit circle, $N_T$ receive signals can be approximately cancelled each other to avoid being intercepted, where $N_T$ is the number of antennas at Alice. Additionally, the subcarrier index spacings from two or more pairs of adjacent transmit antennas should be made to be distinct in order to avoid coherent combining at eavesdropper. Otherwise, they form a energy gathering at eavesdropper with a proper condition. Hence,  $N_T$ subcarrier indices mapped onto the corresponding transmit antennas should be made to be random, distributed, and unequally-spaced.

\item First, we propose two random subcarrier index sets (RSIS): quadratic subcarrier set (QSS) and prime subcarrier set (PSS). Also, linear subcarrier set (LSS) is used as a performance benchmark. Their sizes will directly affect the interception probability of eavesdropper. All three sets may achieve an extremely interception probability by adjusting their parameters. For example, given a fixed value of $N_T$, increasing the RSIS sizes will result in a lower interception probability. In other words, this will achieve an improved anti-interception performance.

\item First, given a RSIS constructed by the above step, $N_T$ subcarriers are randomly chosen from it, and mapped onto the corresponding transmit antennas. To form a sufficiently random mapping from the chosen $N_T$ subcarriers into $N_T$ transmit antennas, a random metric is defined to measure the randomization of the $N_T$ subcarrier indices. Then, a randomization procedure (RP) is proposed to optimize the random metric of the mapping. The total process  includes the following multiple steps:  mod arithmetic  operation, ordering, blocking,  and block interleaving, where block interleaving is repeated until the random metric is larger than a predefined value. Simulation result verifies the fact that the proposed procedure actually generates a high random metric. By  simulation, we find that QSS plus RP and PSS plus RP can form single high energy main peak to achieve a SPWT. It is very difficult for LSS plus RP to implement SPWT. 
\end{enumerate}

The remainder of this paper is organized as follows. Section II describes the system model of SPWT systems with a linear antenna array using RSIS scheme with the aid of AN. In Section III, we give several necessary conditions to reach the goal of SPWT. In Section IV, we firstly constructs two novel RSISs: QSS and PSS. Then, a RP is proposed to achieve a SPWT with SHEMP together with several very weak energy side peaks severely corrupted by AN. In Section V, we make our conclusions.

\section{system model}
\begin{figure}[h]
\centering
\includegraphics[width=0.50\textwidth]{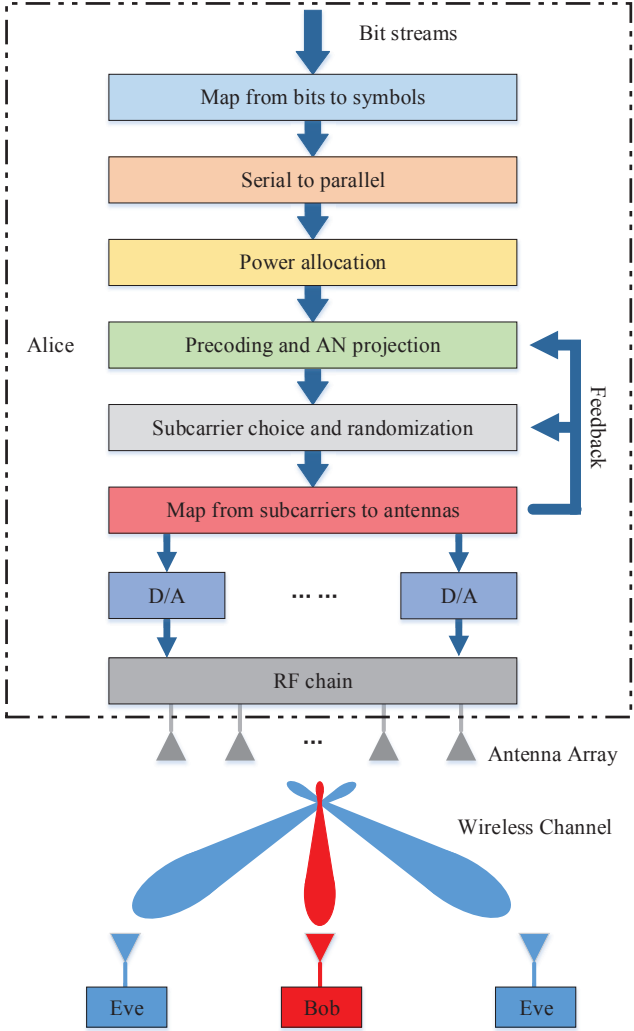}\\
\caption{Block diagram for SPWT systems.}\label{system-model.eps}
\end{figure}

In Fig.\ref{system-model.eps}, a block diagram of SPWT system is shown. Here, Alice is equipped with a $N_T$-element linear antenna array, and all  antennas of Alice transmit the same data symbol per OFDM symbol towards Bob via randomly-selected multiple subcarriers from all-subcarrier set of OFDM. It is assumed that the set of $N_S$ orthogonal subcarriers for OFDM is defined as follows

\begin{equation}\label{f_{sub}}
{S_{sub}} = \left\{ {{f_m}\left| {{f_m} = {f_c} + m\Delta f,\left( {m = 0,1, \ldots N - 1} \right)} \right.} \right\}
\end{equation}
where $f_c$ is the central carrier frequency, $m$ denotes the subcarrier index, i.e., the $m$-th subcarrier, and $\Delta f$ is the subchannel bandwidth. In this paper, we assume that $N_S\Delta f \ll {_c}$, the subcarrier assigned to the $n$-th antenna is ${f_n}$, where ${f_n}\in{S_{sub}}$.
Then, the signal transmitted by antenna  $n$ can be represented by
\begin{align}\label{s_n}
{s_n} &= \sqrt {{\alpha_1 {P_s}}} xe{}^{j {\phi _n} } + \sqrt {{\alpha_2 {P_s}}}  {w_n}
\end{align}
where $x$ is the transmit data symbol with $\mathbb{E}\left\{ {{x^*}x} \right\} = 1$, $w_n$ is the AN, $P_s$ is the total transmit power for all antennas, $\alpha_1$ and $\alpha_2$ are the power allocation (PA) factors which satisfies the identity equation constraint $\alpha _1+\alpha _2=1$, and $\phi _n$ is the initial phase.

In a far-field scenario and line-of-propagation (LoP) channel, the receiver is at arbitrary position $\left({\theta ,R} \right)$, where $\theta$ and $R$ are the angle and distance of the receiver position with respect to the first element of transmit antenna array, respectively. In this work, we choose the first antenna of the array as the reference antenna, then the distance from the $n$-th antenna to receiver can be expressed as ${R_n} = R - \left( {n - 1} \right)d\cos \theta$, where $d$ denotes the element spacing. we define the reference phase as ${\varphi _0}\left( {\theta ,R} \right) =2\pi{f_c}\frac{{{R}}}{c}\ $ and the phase shifting of the $n$-th element as
\begin{align}\label{psi_n}
{\psi _n}\left( {\theta ,R} \right) &= {\rm{ }}{{ 2\pi \left( {{\rm{ }}{f_c} + {k_n}\Delta f} \right){\rm{ }}{{R_n}}/{c}}}-{\varphi _0}\left( {\theta ,R} \right)
\end{align}
So the received signal superimposed from all array elements after down-converted can be given by
\begin{align}\label{Y}
Y\left( {\theta ,R} \right)=\sqrt {\alpha_1 {P_s}}{\mathbf{h}^H}\mathbf{v}x + \sqrt {\alpha_2 {P_s}}{\mathbf{h}^H}\mathbf{w} + n
\end{align}
where
\begin{align}\label{h}
\mathbf{h} =  {\left[ {{e^{j{\psi _1}\left( {\theta ,R} \right)}},{e^{j{\psi _2}\left( {\theta ,R} \right)}}, \ldots ,{e^{j{\psi _{{N_T}}}\left( {\theta ,R} \right)}}} \right]^H},
\end{align}
\begin{align}\label{v}
\mathbf{v} = \frac{1}{{{\sqrt{N_T}}}} {\left[ {{e^{j{\phi _1}}},{e^{j{\phi _2}}}, \ldots ,{e^{j{\phi _{{N_T}}}}}} \right]^H},
\end{align}
and $n$ is the channel noise with the distribution of $n \sim \mathcal{CN}\left( {0,{{\delta}}} \right)$ and $\mathbf{w}$ is the AN.
It is assumed that the desired receiver (Bob) and eavesdroppers (Eve) equipped with  single antenna is located at $\left( {{\theta _B},{R_B}} \right)$, and $\left( {{\theta _E},{R_E}} \right)$, respectively. In general, if Eve is a passive eavesdropper, its position  is unavailable for Alice and Bob. To coherently combine  all $N_T$ signals from $N_T$ antennas at Bob, the initial phase vector $\mathbf{v}$ should be designed to meet the following identity equation
\begin{align}\label{h_B^Hv}
\mathbf{h}_B^H\mathbf{v} = {e^{j{\phi _0}}}
\end{align}
where $\phi_0$ is a constant phase and $\mathbf{h}_B$ is obtained by replacing $\left( {{\theta},{R}} \right)$ with $\left( {{\theta _B},{R_B}} \right)$. In accordance with (\ref{h_B^Hv}), the initial phase of  antenna $n$ should satisfy the following identity
\begin{equation}\label{phi_n}
\phi _n-{\psi _n}\left( {\theta_B ,R_B} \right)=\phi _0,~~~n\in\left\{1,2, \ldots {N_T}\right\}
\end{equation}
In particular, when the desired angle and distance are available, a simple beamforming form $\mathbf{v}_k=\mathbf{h}$ is usually used, thus $\phi_0=0$. Furthermore, the AN vector $\mathbf{w}$ should lie in the null space of $\mathbf{h}_B$, which can be expressed as
\begin{equation}\label{W}
\mathbf{w} = ({\mathbf{I}_{N_T}} - \mathbf{h}_B{\mathbf{h}_B^H})\mathbf{z}
\end{equation}
where $\mathbf{z}$ consists of i.i.d circularly-symmetric complex Gaussian random variables with zero-mean and unit-variance, i.e., $\mathbf{z} \sim \mathcal{CN}\left( {0,{\mathbf{I}_{{N_T}}}}\right)$. Using the null-space projection of AN and matched filter precoder of confidential messages, we have the receive signal at Bob
\begin{align}\label{Y-Rx-Bob}
Y_B\left( {\theta_B ,R_B} \right)=\sqrt {\alpha_1 {P_s}}{\mathbf{h_B}^H}\mathbf{v}x+n_B=\sqrt {\alpha_1 {P_s}}\sqrt{N_T}x+n_B,
\end{align}
which yields the SINR at Bob as follows
\begin{align}\label{SINR_B}
SINR_B = \frac{{{\alpha _1}{P_S}{{\left| {\mathbf{h}_B^H\mathbf{v}} \right|}^2}}}{{{\alpha _2}{P_S}{{\left| {\mathbf{h}_B^H\mathbf{w}} \right|}^2} + \delta^2}}=\frac{{{\alpha _1}{P_S}}}{{{{\delta^2}}}},
\end{align}
In the same manner, we have the receive signal at Eve
\begin{align}\label{Y-Rx-Eve}
Y_E\left( {\theta_E, R_E} \right)=\sqrt {\alpha_1 {P_s}}{\mathbf{h_E}^H}\mathbf{v}x + \sqrt {\alpha_2 {P_s}}{\mathbf{h_E}^H}\mathbf{w} + n_E,
\end{align}
which yields the SINR at Eve as follows
\begin{align}\label{SINR_E}
SINR_E = \frac{{{\alpha _1}{P_S}{{\left| {\mathbf{h}_E^H\mathbf{v}} \right|}^2}}}{{{\alpha _2}{P_S}{{\left| {\mathbf{h}_E^H\mathbf{w}} \right|}^2} + \delta^2}}.
\end{align}
According to (\ref{Y-Rx-Bob}) and (\ref{Y-Rx-Eve}), the original signal $x$ can be readily and successfully detected at Bob due to the array gain from phase alignment. But for Eve, the amplitude and phase of receive signal is seriously distorted and corrupted by AN. In what follows, we will show how to construct a set of random subcarriers and how to perform a RP on the subcarrier indices assigned to all transmit antennas of Alice in order to make all signals cancel each other at Eve. Using a mathematical language, it is preferred that ${\mathbf{h_E}^H}\mathbf{v}\approx 0$ in (\ref{Y-Rx-Eve}) such that Eve has no chance to detect confidential messages successfully.


\section{Several crucial selection rules for designing RSS set}

In this section, we will derive some necessary conditions to achieve a SPWT per OFDM. Our main conclusions are as follows: random, distributed, and unequally-spaced for subcarrier index pattern over transmit antenna array at Alice.

To discuss the SINR at the position of potential eavesdroppers with a given desired user, we need find the correlation between $h_E$ and $h_B$, since $h_E$ is the only variable quantity. First we rewrite (\ref{psi_n}) as
\begin{align}\label{psi_approx}
{\psi _n}\left( {\theta ,R} \right) \approx \frac{{2\pi {k_n}\Delta fR}}{c} - \frac{{2\pi {f_c}\left( {n - 1} \right)d\cos \theta }}{c}
\end{align}
since we have
\begin{equation}\label{2pik_nDelta_f}
\frac{{2\pi {k_n}\Delta f\left( {n - 1} \right)d\cos \theta }}{c} \approx 0
\end{equation}
due to the assumption that $N\Delta f \ll {f_c}$, where array element spacing $d$ is usually set to $d=\lambda/2$ where $\lambda$ is the wavelength.

We find (\ref{SINR_E}) is an increasing function of  $\left|\mathbf{h}_E^H\mathbf{v}\right|$, and
\begin{align}\label{h_e^Hv}
\mathbf{h}_E^H\mathbf{v}=\frac{1}{{{\sqrt{N_T}}}}\sum\limits_{n = 1}^{{N_T}} {{e^{j2\pi \left[ {{k_n}{p_E} - \left( {n - 1} \right){q_E}} \right]}}}
\end{align}
where
\begin{align}\label{p_E}
{p_E} = \frac{{\Delta f({R_B} - {R_E})}}{c},
\end{align}
and
\begin{equation}\label{q_E}
{q_E} = \frac{{{f_c}d\left( {\cos {\theta _B} - \cos {\theta _E}} \right)}}{c}.
\end{equation}

\begin{figure}[h]
\centering
\includegraphics[width=0.50\textwidth]{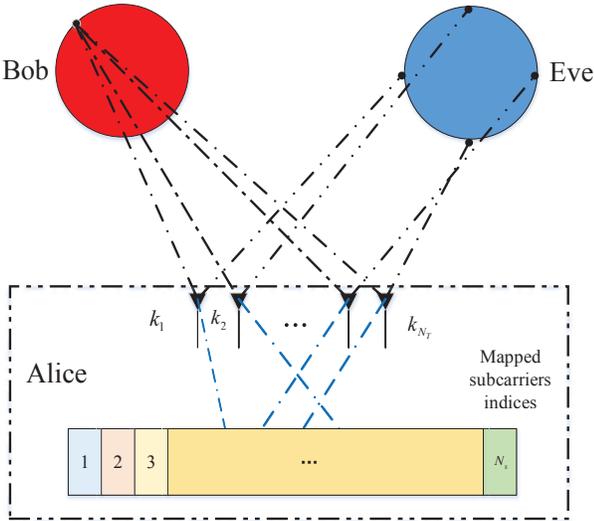}\\
\caption{Schematic diagram of demonstrating the relationship among subcarriers, transmit antennas, and receivers.}\label{mapped.eps}
\end{figure}

We will find that each term in (\ref{h_e^Hv}) is on unit circle in polar coordinate system, as shown in Fig.\ref{mapped.eps}. When all angles of all terms concentrate along one direction or closer, all receive signals from $N_T$ transmit antennas superimpose constructively to form an energy peak. Conversely, if these angles are sparsely,  and uniformly distributed on unit circle,  all receive signals will cancel each other. Finally, the entire receive  signal is such weak that the eavesdropper cannot intercept confidential messages successfully. This will guarantee our proposed SPWT per OFDM.

In a special case that all phases of all terms on the right-side hand of (\ref{h_e^Hv}) are identical,
\begin{align}
2\pi [{{k_n}{p_E} - \left( {n - 1} \right){q_E}}]=\theta_0+2\pi m,
\end{align}
which can be written as
\begin{equation}\label{k_n}
{k_n} = \frac{{\left( {n - 1} \right){q_E} + {\theta _0} + m}}{{{p_E}}}
\end{equation}
where the subcarrier index $k_n$ over antenna $n$ is a linear function of $n$ given an integer value of $m$. In other words, all signals across transmit antenna array is actually viewed as a linear frequency modulation along spatial antenna direction. At this point, all phases align at the undesired receiver. This will forms a strong confidential energy peak, which will yield a serious secure issue. This situation should be avoided.

Now, let us consider a general case, there are two pairs of adjacent antennas and each pair has distinct subcarrier index  spacings, we have
\begin{subequations}
\begin{numcases}{} \label{group_1_1}
   2\pi[{k_{{n_1}}}{p_E} - ({n_1} - 1){q_E}] = 2\pi(\rho + {m_1}), \\ \label{group_1_2}
    2\pi[({k_{{n_1}}}+\Delta k_1){p_E} - {n_1}{q_E}] =2\pi( \rho + {m_2}), \\ \label{group_1_3}
    2\pi[{k_{{n_2}}}{p_E} - ({n_2} - 1){q_E}] = 2\pi(\rho + {m_3}),\\ \label{group_1_4}
    2\pi[({k_{{n_2}}}+\Delta k_2){p_E} - {n_2}{q_E}] =2\pi (\rho + {m_4})
\end{numcases}
\end{subequations}
where $\rho$ is the initial phase with $0 \leq \rho \leq1$, $k_{{n_1}},k_{{n_2}}$ stand for the  subcarrier indices mapped onto antennas $n_1$-th and the $n_2$, and $\Delta k_1$ and $\Delta k_2$ denote the subcarrier index spacings of the two pairs. From the above, by  (\ref{group_1_2}) minus  (\ref{group_1_1}),  (\ref{group_1_3}) minus  (\ref{group_1_2}), and  (\ref{group_1_4}) minus  (\ref{group_1_3}), we have
\begin{subequations}\label{group_2}
\begin{numcases}{} \label{group_2_1}
    \Delta k_1{p_E} - {q_E} =  {m_2}-m_1, \\ \label{group_2_2}
    \Delta k'{p_E} - ({n_2}-n_1 - 1){q_E} = {m_3}-m_2,\\ \label{group_2_3}
    \Delta k_2{p_E} - {q_E} ={m_4}-m_3
\end{numcases}
\end{subequations}
where $\Delta k'={k_{{n_2}}}-k_{{n_1}}-\Delta k_1$, and we can always find a $\rho$ and $m_1$ satisfy equation (\ref{group_1_1}) , and $m_2,m_3,m_4$ also can be adjusted to satisfy equations (\ref{group_2_1}), (\ref{group_2_2}), and (\ref{group_2_3}). Then, let $\Delta k_1=\Delta k_2$, and (\ref{group_2}) may be rewritten as

\begin{subequations}\label{group_3}
\begin{numcases}{} \label{group_3_1}
    \Delta k{p_E} - {q_E} =  m_2-m_1, \\ \label{group_3_2}
    \Delta k'{p_E} - ({n_2}-n_1 - 1){q_E} = m_3-m_2,\\ \label{group_3_3}
    \Delta k{p_E} - {q_E} =m_4-m_3
\end{numcases}
\end{subequations}

From equations (\ref{group_3_1}) and (\ref{group_3_3}), we need $m_4-m_3=m_2-m_1$, thus we rewrite (\ref{group_3}) again as
\begin{subequations}
\begin{numcases}{} \label{group_4_1}
    \Delta k{p_E} - {q_E} =  m_2-m_1, \\ \label{group_4_2}
    \Delta k'{p_E} - ({n_2}-n_1 - 1){q_E} = {m_3}-m_2,\\ \label{group_4_3}
    m_2-m_1=m_4-m_3
\end{numcases}
\end{subequations}
Lastly, from (\ref{group_4_1}) and (\ref{group_4_2}), we can get the solutions to $p_E$ and $q_E$,
\begin{align}\label{p_E_s}
{p_E} = \frac{{\left( {{n_2} - {n_1} - 1} \right)({m_2} - {m_1}) - \left( {{m_3} - {m_2}} \right)}}{{\left( {{n_2} - {n_1} - 1} \right)\Delta k - \Delta k'}}
\end{align}
and
\begin{align}\label{q_E_s}
{q_E} = \frac{{\Delta k'({m_2} - {m_1}) - \Delta k\left( {{m_3} - {m_2}} \right)}}{{\left( {{n_2} - {n_1} - 1} \right)\Delta k - \Delta k'}}
\end{align}
then, from equation (\ref{p_E}) and (\ref{q_E}), we have
\begin{align}\label{R_E_s}
{R_E} = {R_B} - \frac{{{p_E}c}}{{\Delta f}}
\end{align}
and
\begin{align}\label{theta_E_s}
{\theta _E} = \arccos \left( {\cos {\theta _B} - \frac{c}{{{f_c}d}} \cdot {q_E}} \right)
\end{align}
We can get an illegal distance $R_E$ and direction angle $\theta_E$ by substituting (\ref{p_E_s}) and (\ref{q_E_s}) into (\ref{R_E_s}) and (\ref{theta_E_s}), respectively. In such an illegal position, the confidential signals transmitted by the two pairs of adjacent antennas are phase aligned to generate a constructive energy superposition, which means that an energy peak is formed and an about 6B array gain can be achieved. This will enable eavesdropper to intercept easily. Similarly, when there are more pairs of adjacent antennas with the same  subcarrier index spacing, they may have a positive superposition of confidential signal power at some illegal position, which may bring a serious secure problem.

Concluding the above discussion, we have the following theorem

\textbf{Theorem 1:}  To achieve an excellent SPWT per OFDM with single energy main peak and several very weak several side peaks severely corrupted by AN, the corresponding chosen subcarrier indices should be placed along transmit antenna direction at Alice following the three rules: random, distributed, and unequally-spaced.

\section{Proposed two efficient SPWT schemes with randomization procedure}
In this section, we first propose two new RSISs: QSS and PSS with conventional LSS as a performance reference. Then, a RP is proposed to randomize the mapping from subcarriers to transmit antennas by prime modulo operation, and block interleaving. Combining RP with QSS, PSS, and LSS generates three efficient SPWT schemes as follows: QSS plus RP, PSS plus RP, and LSS plus RP. Finally, we derive the closed-form formulas of the three  schemes. 

\subsection{Three methods of constructing RSSs}
Below, three sets of random subcarrier indices are constructed as follows: linear subcarrier set (LSS), quadratic subcarrier set (QSS), and prime subcarrier set (PSS). After this, we can choose $N_T$ subcarriers from one of them and map them onto $N_T $ transmit antennas. The first RSIS is the LSS as follows
\begin{align}\label{linear}
{\mathcal{S}_L} = \left\{ {{k_l}\left| {{k_l} = al + b,k_l\in \mathcal{K}} \right.} \right\}
\end{align}
where $a$ and $b$ are positive integers. The value of $a$ actually has an important impact on the size of set ${\mathcal{S}_L}$. Given a fixed value of $b$, increasing the value of $a
$ will reduce the number of elements in this set. Conversely, decreasing its value will increase the number of elements in this set. The second RSIS QQS is defined as follows
\begin{align}\label{RSS-Set}
{\mathcal{S}_Q} = \left\{ {{k_s}\left| {{k_s} = a{s^2} + bs + c,k_m\in \mathcal{K}} \right.} \right\}
\end{align}
where $a$, $b$, and $c$ are also positive integers. Unlike LSS, the subcarrier index along transmit antenna array  is a non-linear function of the corresponding antenna index.
The final RSIS PSS is defined as
\begin{align}\label{prime}
{\mathcal{S}_P} = \left\{ {{k_p}\left| {k_p~is~a~prime~number,k_p\in \mathcal{K}} \right.} \right\}
\end{align}
where
\begin{align}\label{prime}
{\mathcal{K}} = \left\{ {{k_n}\left| {k_n=n,n=0,1,2,\ldots,N_S-1} \right.} \right\}
\end{align}
where set ${\mathcal{S}_P}$ consists of all primes of not exceeding $N_S-1$. In other words, set ${\mathcal{S}_P}$ stands for all subcarriers with their indices being primes. In general, $N_s$ is far larger than $N_T$.

The above three methods  has a common property that subcarrier indices along antenna direction are in increasing order. This order will yield a receive energy mountain, which will be shown in Section V. This will result in a serious security issue. For example, if eavesdropper moves to any position on this mountain, it will readily intercept confidential messages intended for Bob. How to combat this secure challenging? In the next subsection, we will introduce a RP, which will make its effort to destroy the regular order and make a very random subcarrier index distribution over transmit antennas. This will form a SHEMP at Bob, but in other undesired positions only several weak small energy peaks are immersed in AN energy.

\subsection{Proposed RP}
Given any RSIS shown in Subsection A, for example ${\mathcal{S}_P}$, $N_T$ subcarrier indices are chosen from it to form a  set ${K}$ as
\begin{equation}\label{K}
K=\left\{k_1,k_2,\ldots,k_{N_T}\right\}
\end{equation}
\textbf{Definition 1. Random metric:} To measure the random degree of a random subcarrier set, let us define the following random metric
\begin{equation}\label{delta_I}
\delta_I=\frac{1}{{{N_T} - 1}}{\sum\limits_{i = 1}^{{N_T} - 1} {\left( {{\Delta {{K_i}}} - {{{\bar \Delta} K}} } \right)} ^2}
\end{equation}
where $\Delta {{K_i}}$ is the subcarrier index spacing between a pair of two adjacent antennas $i$ and $i+1$ , and  $\Delta K$ is the  vector of adjacent antenna subcarrier index spacings as follows
\begin{align}\label{I_K}
\Delta K=[\mid K_{T_1}-K_{T_2}\mid,\ldots,\mid K_{T_{N_T-1}}-K_{T_{N_T}}\mid]
\end{align}
where
\begin{equation}\label{barI_K}
{{\bar{\Delta} K}}=\frac{1}{{{N_T} - 1}}\sum\limits_{i = 1}^{{N_T} - 1} {{\Delta {{K_i}}}}
\end{equation}

Observing the definition in $(\ref{delta_I}$), a large $\delta_I$ means that the subcarrier spacing distribution along antenna array is more random, expanded, and distributed. It is very clear that we firstly find a threshold $\delta_{I_T}$. The predefined threshold $\delta_{I_T}$ can be attained by machine learning a large amount of samplings or training data set by randomly choosing $N_T$ subcarriers from $S_P$ or $S_P$. When the subcarrier spacing vector $K$ satisfies
$\delta_{I_f}>\delta_{I_T}$, where $\delta_{I_T}$ is the predefined threshold, it means the $K_T$ is an effective subcarrier index vector, that is, what we need.

\begin{figure}[h]
\centering
\includegraphics[width=0.50\textwidth]{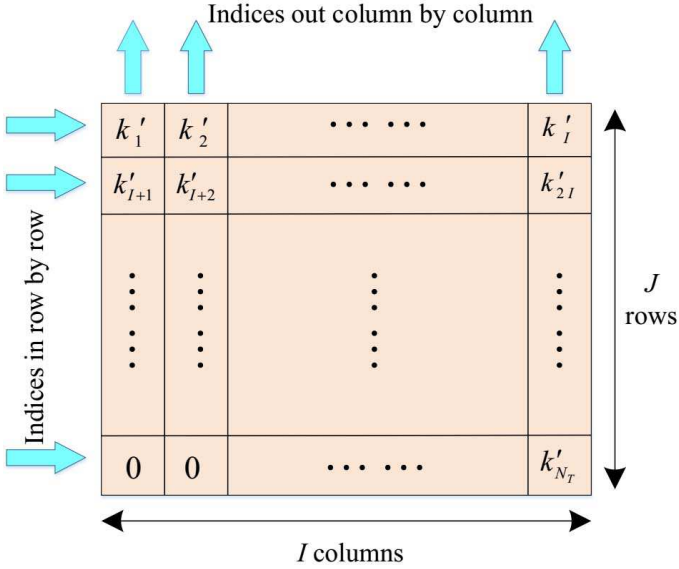}\\
\caption{Block interleaving.}\label{interleaving.eps}
\end{figure}

First, this set is partitioned into $p$ subsets, and represented as
\begin{equation}\label{K'}
{K}'={K}'_0\cup {K}'_1\ldots\cup {K}'_{p-1}
\end{equation}
where ${K}'_n(n=0,\ldots,p-1)$ is a subset with its elements being congruent to be $n$ modulo p, where prime number $p$ is chosen to be a large prime being less than $\sqrt{N_T}$,  that is
\begin{equation}\label{K'_n}
{K}'_n=\{k\in{K}\mid k\equiv n(\mathrm{mod}~p)\},n=0,\ldots,p-1
\end{equation}
where $k\equiv n(\mathrm{mod}~p)$ means that $k$ and $n$ has the same remainder under modulo-p operation.  At the same time, all elements in each subset are arranged in an increasing order. Now, we will permute the element order in the set  $K'$ by block interleaving tool. To implement block interleaving, a $J \times I $ rectangle array is first constructed as shown in Fig.\ref{interleaving.eps}. Then, we place all elements in set $K'$ into this array one by one. Considering that it is possible that $N_T \neq JI$, we sometimes need to add $N_T-JI$ zeros in the front part of the final row of this array such that it is filled completely. Also, the values of $I$ and $J$ are devised to  satisfy the inequality

\begin{equation}\label{N_T}
(J-1)I<N_T<JI
\end{equation}
As shown in Fig.\ref{interleaving.eps}, this scheme inserts zeros to $K'$ and then the following interleaving operation will disorder the sequence of the origin selected subcarrier indices  more efficiently. From Fig.\ref{interleaving.eps}, the $n$th row of the first $N-1$ rows of $A$ is given by
\begin{equation}\label{A_{r_n}}
 A_{r_n}=[k'_{(n-1)I+1},\ldots,k'_{nI}],n=1,2,\ldots,J-1.
\end{equation}
and for convenience, we make the last row with added zeros as
\begin{equation}\label{A_{r_J}}
A_{r_J}=[0,0,\ldots,0,k'_{(J-1)I+1},\ldots,k'_{N_T}].
\end{equation}
Next we get a new subcarrier index sequence $K_T$ after randomization by rewriting $A$ as
\begin{equation}\label{K_T}
K_T=[{A_{c_1}}^T,{A_{c_2}}^T,\ldots,{A_{c_I}}^T]
\end{equation}
where $A_{c_n},n=1,2,\ldots,L$ is the $n$-th column with zero elements removed. This is the zero-padding block interleaving operation. This method can  efficiently randomize the origin sequence so that a sufficient randomization sequence  is formed to achieve a SPWT. In practice, we need to perform several block interleaving operations until the random metric is larger than a predefined threshold to reach a full randomization of subcarriers along antennas.

\subsection{Secure performance analysis}
In Subsection A, three sets, $\mathcal{S}_L$, $\mathcal{S}_Q$, and $\mathcal{S}_P$, of subcarrier indices are presented. Their cardinalities are $M_L$,  $M_Q$, and $M_P$, respectively.

For LSS, the possible ways of choosing $N_T$ from $M_L$ subcarriers is as follows
 \begin{align}\label{LSS-Comb}
C_{{M_L}}^{{N_T}}= \frac{{M_L!}}{{\left( {M_L - {N_T}} \right)!}{N_T!}}
\end{align}
Similarly, we have the possible ways for QSS
\begin{align}\label{QSS-Comb}
C_{{M_Q}}^{{N_T}}= \frac{{M_Q!}}{{\left( {M_Q - {N_T}} \right)!}{N_T!}},
\end{align}
For the convenience of analysis below, $a$, $b$ and $c$ in (\ref{RSS-Set}) are taken to be one, zero and zero, respectively.

Then, the number of elements in QSS is
\begin{align}
M_Q=\lfloor\sqrt{N_S}\rfloor
\end{align}

Substituting the above value in (\ref{QSS-Comb}) yields
\begin{align}\label{QSS-Comb-Simp}
C_{{M_L}}^{{N_T}}= \frac{{\lfloor\sqrt{N_S}\rfloor!}}{{\left( {\lfloor\sqrt{N_S}\rfloor - {N_T}} \right)!}{N_T!}},
\end{align}

With regard to PSS, using the following approximate expression  concerning the number of primes not exceeding $M_P$
\begin{align}\label{M-P}
M_P\approx \frac{{{N_S}}}{{\ln {N_S}}}
\end{align}
in number theory \cite{Graham1998Concrete}, we have the possible ways of PSS
\begin{align}\label{PSS-Comb}
C_{{M_P}}^{{N_T}}= \frac{{M_P!}}{{\left( {M_P-{N_T}} \right)!}{N_T!}}
\end{align}
In general,
\begin{align}\label{M-P-Simp}
M_P\approx \lfloor\ \frac{{{N_S}}}{{\ln {N_S}}}\rfloor>M_Q=\lfloor\sqrt{N_S}\rfloor
\end{align}
for $N_S>2$. Thus, we conclude that the number of combinations or patterns from PSS is greater than that from QSS. For LSS, by adjusting the value of $a$ and $b$, the total number of combinations can be made to be larger or smaller than those of PSS and QSS. In fact, a large value means a more secure transmission. In a practical SPWT network, in order to decrease the interception probability, the desired transmitter transmits confidential messages with a very low power along each subcarrier such that all SNRs  per subcarriers at receivers including desired and undesired are less than 0dB. For an eavesdropper, it is very hard for him or her to decide which subcarriers are used to transmit confidential messages in order to intercept confidential messages. In other words, he should know the transmit subcarrier pattern. However, due to an extremely low receive SNR per subcarrier, Eve only obtains the associated subcarrier pattern by guessing. For Bob, a large transmit antenna array gain is achieved at receiver by SPWT, and can reach a good detection performance. A large number of subcarrier patterns or combinations available imply that she has an extremely small probability to make a successful interception. The probability of interception for LSS is given by
\begin{align}\label{P-L}
P_L=\frac{1}{C_{{M_L}}^{{N_T}}}= \frac{{\left( {M_L - {N_T}} \right)!N_T!}}{{M_L!}}.
\end{align}
In a similar way, we have
\begin{align}\label{P-Q}
P_Q=\frac{1}{C_{{M_Q}}^{{N_T}}}= \frac{{\left( {M_Q - {N_T}} \right)!N_T!}}{{M_Q!}},
\end{align}
and
\begin{align}\label{P-P}
P_P=\frac{1}{C_{{M_P}}^{{N_T}}}= \frac{{\left( {M_P - {N_T}} \right)!N_T!}}{{M_P!}}.
\end{align}
Due to the fact $M_P\gg M_Q$ , as the $N_S$ trend to large, it is very obvious
 \begin{align}\label{P-QandP-P}
P_Q= \frac{{\left( {\lfloor\sqrt{N_S}\rfloor - {N_T}} \right)!N_T!}}{{\lfloor\sqrt{N_S}\rfloor!}}>>P_P=\frac{{ \left( \lfloor{\frac{{{N_S}}}{{\ln {N_S}}} \rfloor - {N_T}} \right)!{N_T}!}}{\lfloor{\frac{{{N_S}}}{{\ln {N_S}}} \rfloor!}}.
\end{align}
However, the value of  $P_L$ is intimately related to the values of $a$ and $b$. Its value may be larger or smaller than those of QSS and PSS. Below, we make an conclusion about the above results as the following theorem:

\textbf{Theorem 2 Interception probability formulas of three RSSs:}  Three random subcarrier sets,  LSS, QSS, and PSS have the following interception probabilities:
\begin{align}\label{P-L}
P_L= \frac{{\left( {M_L - {N_T}} \right)!N_T!}}{{M_L!}},
\end{align}
\begin{align}\label{P-Q}
P_Q= \frac{{\left( {M_Q - {N_T}} \right)!N_T!}}{{M_Q!}},
\end{align}
and
\begin{align}\label{P-P}
P_P= \frac{{\left( {M_P - {N_T}} \right)!N_T!}}{{M_P!}},
\end{align}
respectively.

Observing the three interception probability formulas in Theorem 2, we easily find that the anti-interception performance will be gradually enhanced by increasing the number $N_S$ of total subcarriers provided that the number $N_T$ of antennas at Alice is fixed. This is mainly due to the fact that the values of $M_L$, $M_P$, and  $M_Q$ increase as  $N_S$ increases.

\begin{figure}[h]
\centering
\includegraphics[width=0.53\textwidth]{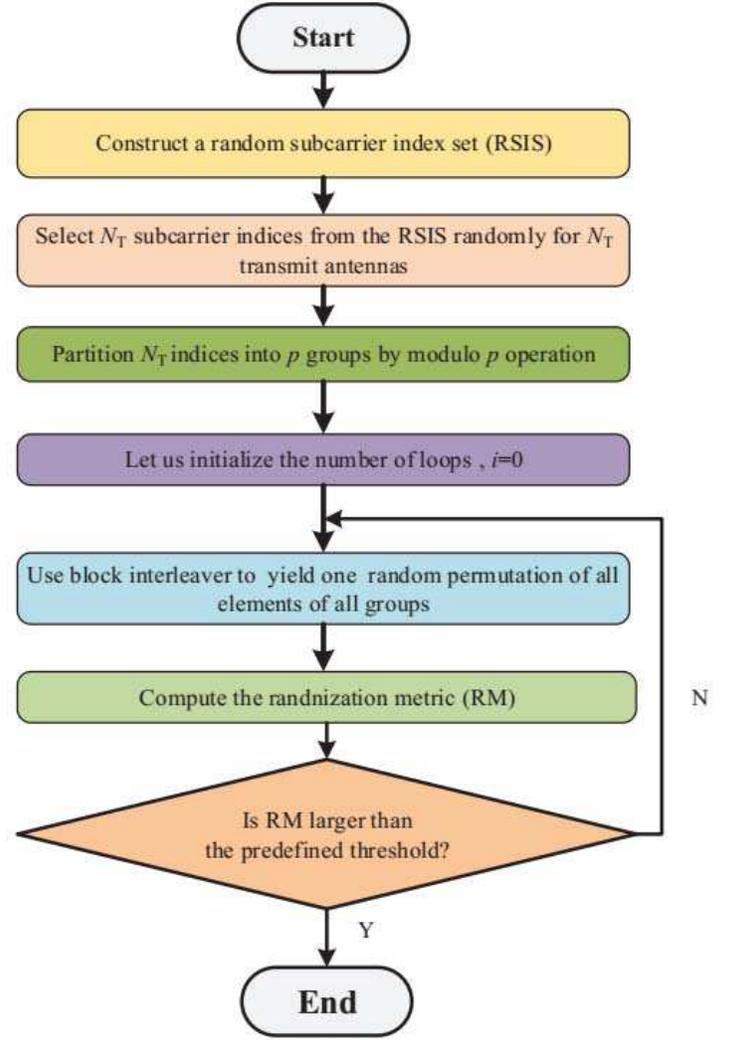}\\
\caption{Block diagram for the proposed RP.}\label{flowchart}
\end{figure}

Finally, we complete this section by presenting a  flowchart as shown in Fig.~\ref{flowchart}. In particular, it is noted that the above three schemes, QSS plus RP, PSS plus RP, and LSS plus RP, follow the steps shown in Fig.~\ref{flowchart}.
\section{Numerical results}
To evaluate the performance of our proposed methods, system parameters in our simulation are chosen as follows: the carrier frequency $f_c=3GHz$, the number of total subcarriers $N=128^2=116384$, $\alpha_1=\alpha_2=0.5$, the number of antenna array element $N_T=120$, $P_s/{\delta}=10dB$, the antenna element spacing $d$ is half of the wavelength (i.e., $d=c/2f_c$), and the position of Bob is  at $(60^{\circ}, 500m)$.

\begin{figure}[h]
\centering
\includegraphics[width=0.50\textwidth]{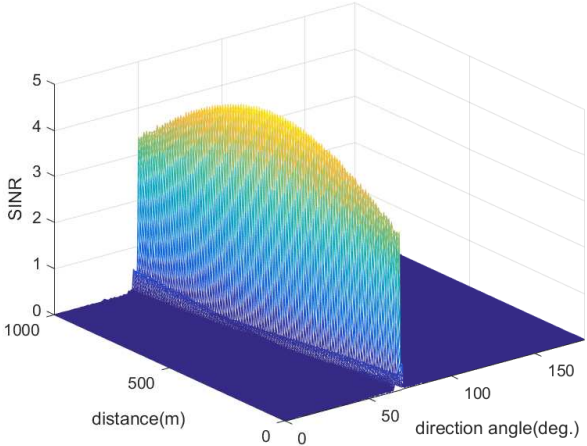}\\
\caption{3-D surface of SINR versus direction angle and distance of LSS without RP.}\label{linear.eps}
\end{figure}

Fig.\ref{linear.eps} illustrates the 3-D performance surface of SINR versus direction angle $\theta$ and distance $R$ of LSS without RP.  From this figure, it is seen that the SINR of LSS will have a high  energy ridge or mountain chain extending to infinity. If eavesdropper locates on this ridge, then it can intercept confidential messages. This will generate a serious security issue. Thus, this LSS scheme without RP should be avoided in practical applications.

\begin{figure}[h]
\centering
\includegraphics[width=0.50\textwidth]{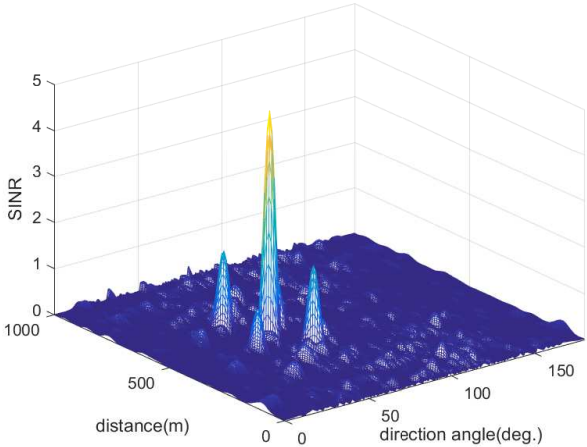}\\
\caption{3-D surface of SINR versus direction angle and distance of LSS plus RP.}\label{linear_interleaving.eps}
\end{figure}

To solve the secure problem of the LSS without RP in Fig.\ref{linear.eps}, we combine LSS and RP to form LSS plus RP.   Fig.\ref{linear_interleaving.eps} shows the 3-D performance surface of SINR versus direction angle $\theta$ and distance $R$ of LSS plus RP. From this figure,  it is seen that  the secure performance are significantly improved, but there still are two lower energy side peaks outside the desired main peak, which will still lead to a secure issue if the eavesdropper locates on the two lower side peaks, and can still receive a good quality of confidential signal. In other words, the LSS plus RP cannot completely solve the secure issue. Below, we will turn to QSS and PSS.

\begin{figure}[h]
\centering
\includegraphics[width=0.50\textwidth]{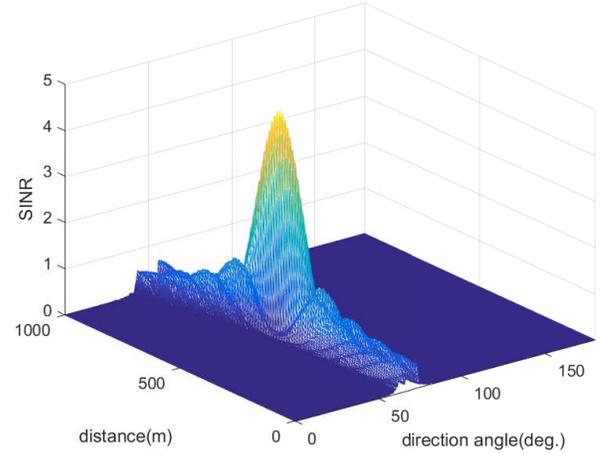}\\
\caption{3-D surface of SINR versus direction angle and distance of the proposed QSS without RP.}\label{square.eps}
\end{figure}

Fig.\ref{square.eps} plots the 3-D performance surface of SINR versus direction angle $\theta$ and distance $R$ of  QSS without RP.  From this figure, it is seen that the SINR of QSS will also have a high energy ridge or mountain chain extending to infinity although it performs better than  LSS. This will also generate a security issue. Thus, this QSS  without RP should also be avoided in practical applications.

\begin{figure}[h]
\centering
\includegraphics[width=0.50\textwidth]{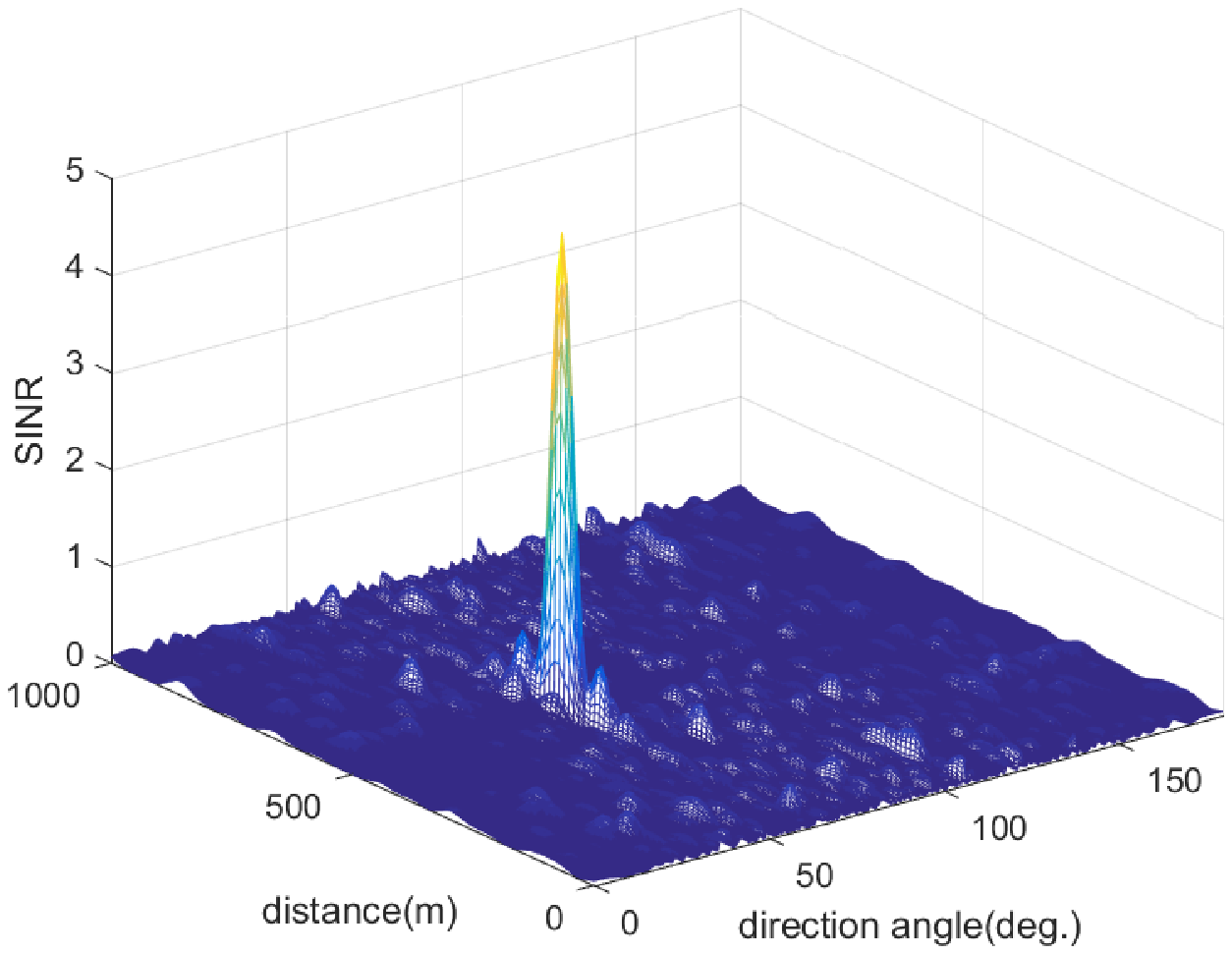}\\
\caption{3-D surface of SINR versus direction angle and distance of the proposed QSS plus RP.}\label{square_interleaving.eps}
\end{figure}

Similar to LSS, we also introduce RP to QSS, Fig.\ref{square_interleaving.eps} plots the 3-D performance surface of SINR versus direction angle $\theta$ and distance $R$ of  QSS plus RP. After RP, compared with Fig.\ref{square.eps},  the secure performance have been improved greatly, and there exists only several very weak energy side peaks outside the energy main peak. Compared to the main peak, their energy can be omitted. The best value of side peaks is about one sixth of the main peak. Obviously, the QSS plus RP can achieve a SPWT.

\begin{figure}[h]
\centering
\includegraphics[width=0.50\textwidth]{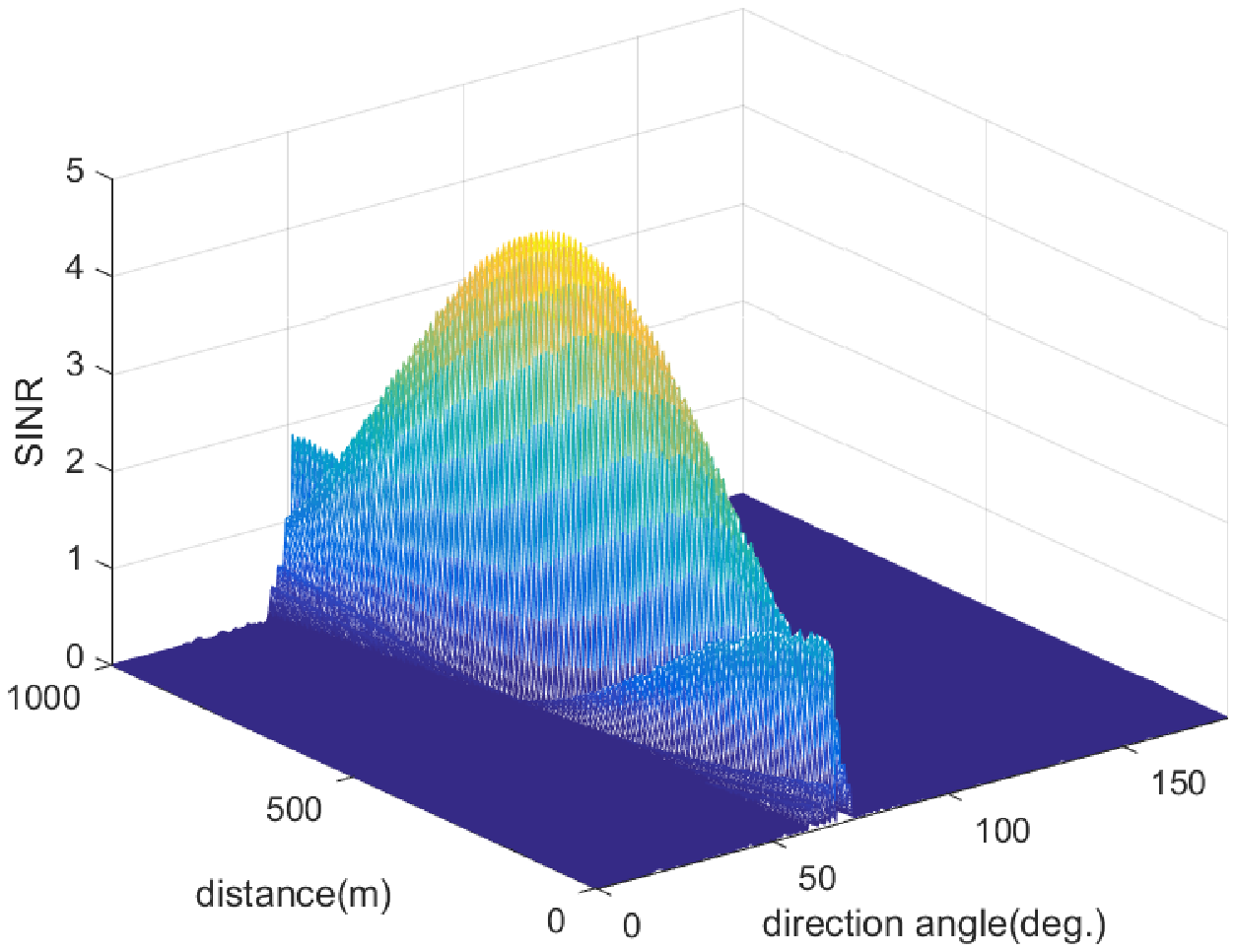}\\
\caption{3-D surface of SINR versus direction angle and distance of the proposed PSS without RP.}\label{prime.eps}
\end{figure}

\begin{figure}[h]
\centering
\includegraphics[width=0.50\textwidth]{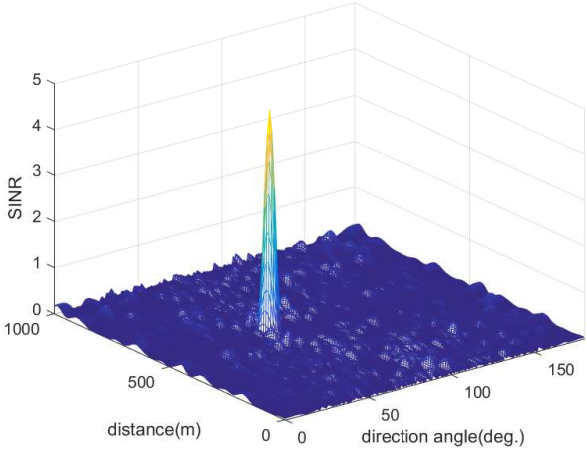}\\
\caption{3-D surface of SINR versus direction angle and distance of the proposed PSS plus RP.}\label{prime_interleaving.eps}
\end{figure}

%
Now, we move to PSS. Fig. \ref{prime.eps} demonstrates the 3-D SINR performance surface of PSS without RP. Obviously,  the same secure problem also exists in PSS without RP. Similar to LSS and QSS, it also has a high energy ridge or mountain chain. To combat this secure problem, Fig. \ref{prime_interleaving.eps} demonstrates the 3-D SINR performance surface of PSS plus RP. Observing this figure,  it is seen that there is only one high signal energy peak of confidential messages formed at desired position. Several weak side peaks are also presented outside the main peak, but their peak values  are far lower than that of the main peak. By our coarse measurement, their peak values are only one tenth of the main peak value. 
%
%
%

\begin{figure}[h]
\centering
\includegraphics[width=0.50\textwidth]{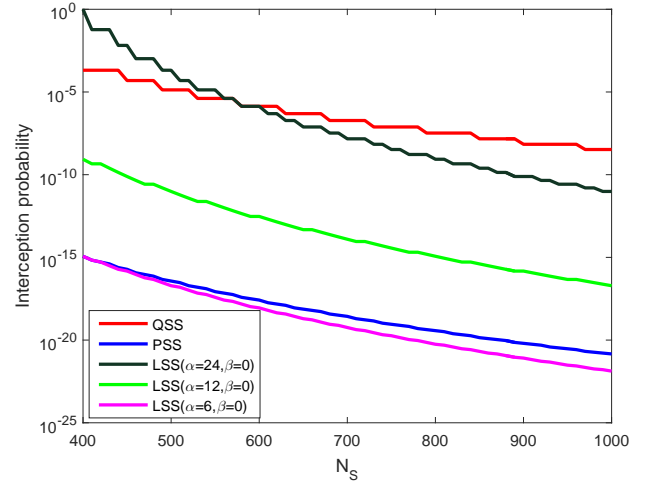}\\
\caption{Curves of interception probability versus $N_S$.}\label{PN_S}
\end{figure}

\begin{figure}[h]
\centering
\includegraphics[width=0.50\textwidth]{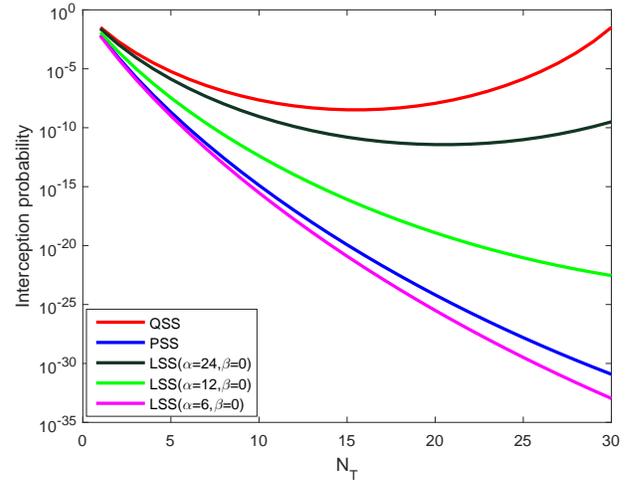}\\
\caption{Curves of interception probability versus $N_T$.}\label{PN_T}
\end{figure}

In what follows, we will make a comparison of their interception probabilities. Fig.\ref{PN_S} demonstrates the curves of intercept probability versus the number $N_S$ of total subcarriers  of LSS plus RP, QSS plus RP, and PSS plus RP with $N_T=16$. We can find that their interception probabilities decrease with  increasing the value of $N_S$. The interception probability of QSS plus RP is the worse one among the three schemes due to the fact that the number of subcarrier indices in QSS are  smaller than those of LSS and PSS. Furthermore, with regard to LSS, reducing the subcarrier spacing between two adjacent antennas will increase the number of possible chosen subcarriers. This will also increase the size of LSS, and accordingly decrease the interception probability. In other words, the anti-interception performance is improved.

Fig.\ref{PN_T} illustrates the curves of interception probability versus the number $N_T$  of antennas  of LSS, QSS, and PSS with $N_S=1000$. It is shown that  as  $N_T$ increases, the  interception probability of  LSS and PSS has improved gradually, but the interception probability  of QSS is a convex function of  $N_T$    due to the fact that the number of total subcarriers is limited and increasing the value of $N_T$ will reduce the possible selectable subcarrier patterns.  At the same time, the anti-interception performance of LSS  improves as the subcarrier spacing of LSS decreases. In summary, the LSS can achieve an excellent anti-interception performance provided that  the subcarrier spacing is chosen to be small enough. In general, PSS is in between LSS and QSS. QSS is the worst anti-interception performance.

In summary, from Fig.~5 to Fig.~10, we find some  basic facts as follows.  QSS plus RP and PSS plus RP have a less energy leakage compared with LSS plus RP. Via RP, there still exists two high energy side peaks for the LSS plus RP. For  QSS  and PSS, after RP operation, outside the main peaks, there are several very weak energy peaks, which can be neglected compared to the main peaks. The most important fact is that RP actually converts confidential energy mountain into a high confidential energy main peak, and reduces energy leakage. Observing Figs. 11 and 12, we also find that the interception probability of PSS plus RP is between QSS plus RP and LSS plus RP. Considering the two faces, it is evident that the proposed PSS plus RP make a good balance between SPWT and anti-interception performance. The proposed schemes  QSS plus RP and PSS plus RP will pave a way for the near future applications of SPWT.

\section{Conclusions}
In this paper, to achieve a practical SPWT, two SPWT schemes are proposed: QSS plus RP, and PSS plus RP. First, we derived some necessary conditions for SPWT by number theory. From analysis and derivation, it follows that  a random, distributed, and  non-equidistant subcarrier pattern placed on transmit antenna array is preferred. Then, three random subcarrier sets are presented: LSS, QSS, and PSS. The later two sets are proposed by us. At the same time, the random metric is defined to measure the degree of randomization of subcarrier pattern along antenna direction. Following this, a RP scheme is proposed to implement a true SPWT. The RP includes modulo operation, ordering, and block interleaving. From the analysis and simulations, our proposed two methods perform better than LSS plus RP in terms of security, where security implies a less energy leakage and a formed single energy main peak with several weak side peaks omitted. By analysis and simulation, we also find that the proposed PSS plus RP has a better anti-interception performance compared to QSS plus RP. Although the LSS plus RP can achieve the lowest interception probability among the three schemes by reducing the subcarrier spacing, it will result in a serious energy leakage outside the main peak, which yields a serious secure issue. In summary, it is very obvious that the proposed PSS plus RP can strike a good balance between energy leakage and anti-interception performance. The proposed SPWT schemes provide  an attractive alternative secure solution to the future satellite communications, unmanned aerial vehicle,  and 5G and beyond networks.

\ifCLASSOPTIONcaptionsoff
  \newpage
\fi

\bibliographystyle{IEEEtran}
\bibliography{IEEEfull,reference}
\end{document}